\documentclass[12pt]{article}
\usepackage{cite}
\usepackage{amsmath, amsthm, amssymb,slashed}
\usepackage{ifpdf}
\ifpdf
  \usepackage[pdftex]{graphicx}
  \usepackage{epstopdf}
\else
  \usepackage[dvips]{graphicx}
\fi
\parskip=\baselineskip

\def\Tr{{\rm Tr}}
\def\16{{\bf 16}}
\def\1{{\bf 1}}
\def\2{{\bf 2}}
\def\3{{\bf 3}}
\def\4{{\bf 4}}

\def\hat{\widehat}
\font\teneurm=eurm10 \font\seveneurm=eurm7 \font\fiveeurm=eurm5
\newfam\eurmfam
\textfont\eurmfam=\teneurm \scriptfont\eurmfam=\seveneurm
\scriptscriptfont\eurmfam=\fiveeurm

\font\teneusm=eusm10 \font\seveneusm=eusm7 \font\fiveeusm=eusm5
\newfam\eusmfam
\textfont\eusmfam=\teneusm \scriptfont\eusmfam=\seveneusm
\scriptscriptfont\eusmfam=\fiveeusm

\font\tencmmib=cmmib10 \skewchar\tencmmib='177
\font\sevencmmib=cmmib7 \skewchar\sevencmmib='177
\font\fivecmmib=cmmib5 \skewchar\fivecmmib='177
\newfam\cmmibfam
\textfont\cmmibfam=\tencmmib \scriptfont\cmmibfam=\sevencmmib
\scriptscriptfont\cmmibfam=\fivecmmib

\numberwithin{equation}{section}

\def\be{\begin{equation}}
\def\ee{\end{equation}}

\begin{document}
\begin{titlepage}
\begin{flushright}
\end{flushright}

\vskip 1in
\begin{center}
{\bf\Large{ Axial-Anomaly in Noncommutative QED  and \vskip3mm Pauli-Villars Regularization}}
\vskip
1cm\centerline {Mohammad Walid AlMasri }\vskip .05in
{\small\textit{Physics Department, Ko\c{c} University, 
Sarıyer 34450,Istanbul,Turkey}}\\ Emails: \texttt{mmasri14@ku.edu.tr, mwalmasri2003@gmail.com}
\end{center}
\vskip 0.5in
\centerline{\bf{Abstract}}
\vskip.05in{We calculate the $U(1)$ axial-anomaly in two and four dimensions using a modified path integral method coupled to a Pauli-Villars regulator field in the noncommutative QED. Pauli-Villars regularization method provides us with unambiguous way to connect the modified path integral formalism with perturbative axial Ward identities at each step of calculations. }
\baselineskip 16pt

\end{titlepage}

\tableofcontents
\section{Introduction}\label{intro}

Quantum anomaly occurs when the conservation laws are valid in the classical level but it is violated
at the quantized level.
One basic feature of quantum field theory is the principle of gauge symmetry. Its breakdown violates the corresponding conserved current.
For example, in some theories of fermions with chiral symmetry, the quantization may
lead to the breaking of chiral symmetry which in turn implies a non-conserved chiral
charges.
The axial anomaly was first studied by Jack Steinberger in  1949 when he
calculated the Feynman Triangle diagram in ``Pion-Nucleon'' $\pi$-N  Model which
contains a $\gamma_{5}$ vertex.
in 1963, Johnson pointed out that in a massless 2-dimensional QED one cannot have
both the conservation of gauge current and axial current. It was proved by
performing the Feynman triangle diagrams which made up of one-axial current and
two-vector currents containing an UV-divergence. They found that conservation of vector current implies the breaking of conversation law for axial current.  
In 1969 Adler-Bardeen showed that anomaly is already given by 1-Loop
calculations. Thus, axial anomaly receives no contributions from radiative
corrections.
This means that due to loop corrections, the symmetry might be
spoiled and this symmetry will not be the symmetry of effective action. \\
Non-commutative quantum field theory has been studied extensively in the last two decades \cite{Connes,Seiberg,Minwalla}. It has been  shown that an UV/IR mixing arises in noncommutative theories. 
It also emerges in the context of  perturbative string theory in the presence of D-branes with non-vanishing $B$-Field background \cite{Ardalan,Lledo}. \\
The $U(1)$ axial anomaly in the non-commutative QED has been calculated  and it is well understood, 
\cite{Sadooghi,Martin}. In this work, We will calculate the $U(1)$ axial anomaly in two and four dimensions using Pauli-Villars regularization method.

The layout of this paper as follows, In first section of this work, we shall give a brief review on Non-commutative $U(1)$ gauge theory. In the second section, we will calculate the axial anomaly in two and four dimensions using our modified path integral coupled to Pauli-Villars heavy fields. At the end, we  conclude and give some advantages of our method.
The paper is followed by a an appendix  Fujikawa's method  in calculating axial anomaly in NC-QED.

\section{Noncommutative $U(1)$ Gauge Theory}
Non-commutative field  theories are defined on space-time which the following commutation relation between 
its coordinates is satisfied $ [\hat{x}_{i},\hat{x}_{j}]=i\;\theta^{ij}$, where $\theta^{ij}$ is a $\mathbb{R}^{4}\bigotimes\mathbb{R}^{4}$
real antisymmetric constant. The usual product is promoted to $\star$-Moyal product in the non-commutative theories.  Let $C^{\infty}(\mathbb{R}^{2n})$ be the space of smooth functions, 
$f:\mathbb{R}^{2n}\rightarrow \mathbb{C}^{n}$, for $f,g$  $\epsilon$ $C^{\infty}(\mathbb{R}^{2n})$
the Moyal $\star$ product is defined as 
\begin{equation}
f\;\star\; g(x)= e^{i\frac{\theta_{\mu\nu}}{2}\; \frac{\partial}{\partial{\zeta_{\mu}}}\; \frac{\partial}{\partial{\eta_{\nu}}}}\; f(x+\zeta)\; g(x+\eta) \mid_{\zeta=\eta=0}
\end{equation}
Where $\theta_{\mu\nu}$ is a real antisymmetric constant.The previous Moyal Star $\star$ product satisfies the $C^{\star}$-algebra.  by expanding the previous formula up to first order, we find that 
\begin{equation}
f\;\star g= f\;q+\frac{i}{2}\; \theta^{\mu\nu}\; \partial_{\mu}f\; \partial_{\nu}g+ \mathcal{O}(\theta^{2})
\end{equation}
The star product has many properties like \\
\begin{equation}
f\; \star g \mid _{\theta}=g\; \star f \mid_{-\theta}
\end{equation}
Suppose $h$ is another smooth function , we have the following cyclic property 
\begin{equation} \label{cyclic}
\int_{-\infty}^{\infty}\; d^{D}x\; ( f \star g \star h)=\int_{-\infty}^{\infty}\; d^{D}x\; ( h \star f \star g)=\int_{-\infty}^{\infty}\; d^{D}x\; ( g \star h \star f)
\end{equation}
The regular commutator is generalized in the non-commutative case to the Moyal bracket
\begin{equation}
[f,g]_{\star}=f\star\;g - g\star\;f
\end{equation}
The deformed Jacobi identity is 
\begin{equation}
[f,[g,h]_{\star}]_{\star}+[h,[f,g]_{\star}]_{\star}+[g,[h,f]_{\star}]_{\star}=0
\end{equation}
Remarkably, the commutator in Moyal product involves both ordinary commutators and anti-commutators\cite{Chams}.
\begin{equation}
f\star\;g - g\star\;f = [f,g]_{(\star,even)}+ \{f,g\}_{(\star,odd)}
\end{equation}
\begin{equation}
[f,g]_{(\star,even)}=[f,g]+(\frac{i}{2})^{2}\; \theta^{\mu\nu}\theta^{\rho\sigma}[\partial_{\mu}\partial_{\rho}f,\partial_{\nu}\partial_{\sigma}g]+\mathcal{O}(\theta^{4})
\end{equation}
\begin{equation}
\{f,g\}_{(\star,odd)}=\frac{i}{2}\theta^{\mu\nu}\{\partial_{\mu}f,\partial_{\nu}g\}+(\frac{i}{2})^{3}\theta^{\mu\nu}\theta^{\rho\sigma}\theta^{\kappa\lambda} \{\partial_{\mu}\partial_{\rho}\partial_{\kappa}f,\partial_{\nu}\partial_{\sigma}\partial_{\lambda}g\}+\mathcal{O}(\theta^{5})
\end{equation}
Let us now define symmetric and antisymmetric star product as 

\begin{equation}
(f\star g)_{s}=\frac{1}{2}(f\star g+ g \star f)= fg+(\frac{i}{2})^{2}\; \theta^{\mu\nu}\theta^{\rho \sigma}\; \partial_{\mu}\partial_{\rho}f\; \partial_{\nu}\partial_{\sigma} g+\mathcal{O}(\theta^{4})
\end{equation}

\begin{equation}
(f \star g)_{a}=\frac{1}{2}(f \star g-g \star f)= (\frac{i}{2}) \theta^{\mu\nu} \; \partial_{\mu} f \partial_{\nu} g+ (\frac{i}{2})^{3}\theta^{\mu\nu}\theta^{\rho\sigma}\theta^{\kappa \lambda} \; \partial_{\mu}\partial_{\rho}\partial_{\kappa}f \; \partial_{\nu}\partial_{\sigma}\partial_{\lambda}g
+\mathcal{O}(\theta^{5})
\end{equation}
The covariant derivative can be written as  
\begin{equation}
D_{\mu}=\partial_{\mu}+i\;g\;[A_{\mu}, \; \; ]_{\star} 
\end{equation}
Where $A_{\mu}$ is the $U(1)$ gauge connection. The  two-form field strength tensor is given by the commutator 
\begin{equation}
[D_{\mu},D_{\nu}]_{\star}=i\;g\; F_{\mu\nu}(x)
\end{equation}
\begin{equation} \label{fieldstrength}
F_{\mu\nu}(x)=\partial_{\mu}A_{\nu}(x)-\partial_{\nu}A_{\mu}(x)+ig [A_{\mu}(x),A_{\nu}(x)]_{\star}
\end{equation}
The field strength $ F_{\mu\nu}$ transforms like 
\begin{equation}
F_{\mu\nu}(x) \rightarrow F_{\mu\nu}^{'}(x)=U(x)\star \; F_{\mu\nu}(x)\;  \star U^{-1}(x)
\end{equation}
From the equation \eqref{fieldstrength} we see that the two-form field strength tensor has a similar form to field strength in Yang-Mills gauge theory, so noncommutative QED exhibits properties like the usual Yang-Mills theories which has non-Abelian gauge group at the commutative level. \\
The pure ( not-coupled to matter) non-commutative $U(1)$ gauge action in any arbitrary dimension $D$ is 
\begin{equation}
S_{\emph{gauge}}[A_{\mu}]=-\frac{1}{4}\; \int d^{D}x\; F_{\mu\nu}(x)\star F^{\mu\nu}(x)
\end{equation}

The non-commutative gauge action transforms as 
\begin{equation}
S_{\emph{gauge}}\rightarrow S^{'}_{\emph{gauge}}=-\frac{1}{4}\; \int d^{D}x\; U(x)\star \; F_{\mu\nu}\star \; U(x)^{-1}\star \; U(x)\star \; F^{\mu\nu}(x)\star \; U(x)^{-1}
\end{equation}

Note that the $U(1)$ Lie group is deformed in the non-commutative case. It can be expanded as \cite{Lledo}
\begin{align}
U(x)_{\star}=(e^{i\;a})_{\star}=\sum_{n=o}^{\infty}\; \frac{(ia)^{n}_{\star}}{n!}
\end{align}
On other hand ,the unitarity condition still valid in the non-commutative  as it should be since we are seeking a unitary class of transformations 
\begin{equation}
U(x)\star \; U^{-1}(x)=U^{-1}(x)\star \; U(x)=\mathbb{I}
\end{equation}
Where $ \mathbb{I}$ is the $N\; \times \; N$  identity matrix , where $N$ is the dimension of  the Lie group representation. 
In the fundamental representation of $U(1)$, the identity matrix $ \mathbb{I}$ is simply  the number one.\\ 
The local gauge transformations of the gauge field are  
\begin{equation}
A_{\mu}\rightarrow A^{'}_{\mu}(x)=U(x)\star A_{\mu}\star U^{-1}(x)+\frac{i}{g}(\partial_{\mu}U)\star U^{-1}(x)
\end{equation}
where $U$ is the unitary transformation matrix.it  has the form $U(x)=(e^{ig\;a(x)})_{\star}$ and it can be expanded as follows 
\begin{equation}
U(x)=(e^{ig\;a(x)})_{\star}=1+ig\; a(x)-\frac{g^{2}}{2}a(x)\star a(x)+\mathcal{O}(a^{3})
\end{equation}
The Fermionic (matter) action can be expressed as
\begin{equation}
S_{F}[\overline{\psi},\psi]=\int d^{D}x \; [\;i\overline{\psi}(x)\;\gamma^{\mu}\star\; D_{\mu}\; \psi(x)-m\; \overline{\psi}(x)\;\star \psi(x)\;]
\end{equation}
The covariant derivative that acts on $\psi$ is
\begin{equation}
D_{\mu}\psi=\partial_{\mu}\psi(x)+ig\;A_{\mu}\star \psi(x)
\end{equation}
The transformation of matter fields are 
\begin{align}
\psi^{'}(x)=U(x)\star \psi (x), \\ \nonumber
\overline{\psi^{'}}(x)=\overline{\psi}(x)\star U^{-1}(x).
\end{align}
Up to first order in parameter $a(x)$, the previous matter fields transform like
\begin{align}
\psi^{'}(x)\rightarrow \psi(x)+ig\;a(x)\star \psi(x), \\ \nonumber
\overline{\psi^{'}}(x)\rightarrow \overline{\psi}(x)-ig\; \overline{\psi}(x)\star a(x). 
\end{align}
In the noncommutative case, the conserved current is 
\begin{equation}
j^{\mu}=\overline{\psi}(x)\star \gamma^{\mu}\psi(x)=\overline{\psi}\gamma^{\mu}\psi+\frac{i}{2}\; \theta^{\mu\nu}\; \partial_{\mu}\overline{\psi}\gamma^{\mu}\partial_{\nu}\psi+\dots 
\end{equation}

Finally, we can write the total NC-QED action after considering the matter+gauge+ghost contributions altogether as  
\begin{align}\label{qed}
S_{\emph{NC-QED}}[A_{\mu},\overline{\psi},\psi]=\int d^{D}x\; -\frac{1}{4}\;  F_{\mu\nu}(x)\star F^{\mu\nu}(x)+ \overline{\psi}(x)\star (i\slashed{D}-m)\; \psi(x)\\ \nonumber +\mathcal{L}_{\emph{gauge-fixing}}+\mathcal{L}_{\emph{ghost}}
\end{align}
Where the gauge-fixing action $ S_{\emph{gauge-fixing}}$ and the ghost action $S_{\emph{ghost}}$ are given respectively as 
\begin{equation}
S_{\emph{gauge-fixing}}[A_{\mu}]=-\frac{1}{2\xi}\int d^{D}x \;  (\partial_{\mu}A_{\mu})^{2} , 
\end{equation}

\begin{equation}
S_{\emph{ghost}}[A_{\mu},C,\overline{C}]=\int d^{D}x \;  \partial_{\mu}\overline{C}(\partial_{\mu}C+ig[A_{\mu},C]_{\star}). 
\end{equation}
Where $C$ and $\overline{C}$ are the Faddeev-Popov ghosts. 

The  corresponding Euler-Lagrange equations for the total NC-QED \ref{qed} are
\begin{equation}
i\gamma^{\mu}\partial_{\mu}\psi-\;m\psi-g\gamma^{\mu}A_{\mu}\star \psi=0 ,
\end{equation}
\begin{equation}
i\partial_{\mu}\overline{\psi}\gamma^{\mu}+m \overline{\psi}+g \overline{\psi} \gamma^{\mu}\star A_{\mu}=0 ,
\end{equation}
\begin{equation}
D_{\mu}F^{\mu\nu}+\frac{1}{\xi}\partial^{\nu}\partial_{\mu}A^{\mu}=gj^{\nu}. 
\end{equation}

\section{Pauli-Villars Regularization Method}
The main idea of Pauli-Villars regularization scheme is to introduce a heavy fields  (regulators) and sending their masses  
to infinity at the end of calculations. Thus integrals of the form 
\begin{equation}
\mathcal{I}=\int \frac{d^{4}k}{(2\pi)^{4}} \; \frac{1}{(k^{2}-m^{2}+i\epsilon)}
\end{equation}
will become in the Pauli-Villars regularization scheme like
\begin{equation}
\mathcal{I}_{\emph{Pauli-Villars scheme }}=\int \frac{d^{4}k}{(2\pi)^{4}} \big[\; \frac{1}{(k^{2}-m^{2}+i\epsilon)}- \frac{1}{(k^{2}-M^{2}+i\epsilon)} \big]
\end{equation}
Where $k$ is the momentum of the particle, $m$ its mass and  $M$ is the fictitious large mass which corresponds to Pauli-Villars field
which are usually referred to as  $\emph{ghost}$. \\ 

The calculation of chiral anomaly in the commutative  gauge theories using Pauli-Villars method was elaborated in \cite{Fujikawa,Banerjee}. 
\subsection{Two-Dimensional Anomaly}
The modified NC - massless Dirac action in two-dimensions is 
\begin{equation}
S_{F}=\int d^{2}x\; [ \overline{\psi}(x)\star\; i\gamma^{\mu}D_{\mu}\;\star  \psi(x)+ \overline{\eta}(x)\star \; (i\gamma^{\mu}D_{\mu}-M)\;\star  \eta(x)]
\end{equation}
Where $\psi(x)$, $\overline{\psi}(x)$ are the two-dimensional spinor fields and $\eta(x)$, $\overline{\eta}(x)$ 
are Bose-like (unphysical ) spinor fields. These unphysical fields are represented as complex numbers rather than Grassmann numbers like the usual physical spinors.
We adopt Pauli  matrices as basis throughout our calculations of the  axial anomaly in  two dimensions
\begin{eqnarray}\sigma_{1}=\gamma_{1}=\left(
\begin{array}{ c c }
0& 1 \\
1 & 0
\end{array} \right),\;\;\; 
\sigma_{2}=\gamma_{2}=\left(
\begin{array}{ c c }
0& -i \\
i & 0
\end{array} \right),\;\;\;
\sigma_{3}=\gamma_{3}=\left(
\begin{array}{ c c }
1& 0\\
0 & -1
\end{array} \right)
\end{eqnarray}
$\gamma_{3}$ is the chirality operator and it is defined to be $\gamma_{3}=-i\; \gamma_{1}\gamma_{2}$. \\
The trace formulas in two-dimensions are 
\begin{equation}
\Tr(\gamma_{\mu}\gamma_{\nu})=2g_{\mu\nu}
\end{equation}
\begin{equation}
\Tr(\gamma_{3}\gamma_{\mu}\gamma_{\nu})=2i\; \epsilon_{\mu\nu}
\end{equation}
The Euclidean effective action is 

\begin{equation}
e^{-W}=\int \mathcal{D}\psi \; \mathcal{D}\overline{\psi}\; \mathcal{D}\eta\; \mathcal{D}\overline{\eta}\; e^{-S_{F}} ,
\end{equation}
and the corresponding  conserved vector   current is 
\begin{equation}
J_{\mu}=\overline{\psi}(x)\star\; \gamma_{\mu}\; \psi(x)+\overline{\eta}(x)\star\; \gamma_{\mu}\; \eta(x). 
\end{equation}
The associated axial-vector current is 
\begin{equation}
J^{3}_{\mu}=\overline{\psi}(x)\star\; \gamma_{\mu}\gamma^{3}\; \psi(x)+\overline{\eta}(x)\star\; \gamma_{\mu}\gamma^{3}\; \eta(x) .
\end{equation}
Under chiral transformation and up to first order in $a(x)$, the matter fields transform like 
\begin{align}
\psi^{'}(x)\rightarrow  \psi(x)+i\; \gamma_{3}\; a(x)\; \star \psi(x), \\ \nonumber
\overline{\psi^{'}}(x)\rightarrow \overline{\psi}(x)+i\; \overline{\psi}(x)\star \; \gamma_{3}\; a(x), \\ \nonumber
\eta^{'}(x)\rightarrow \eta(x)+i\; \gamma_{3}\; a(x)\star \; \eta(x), \\ \nonumber
\overline{\eta^{'}}(x)\rightarrow \overline{\eta}(x)+i\; \overline{\eta}(x)\star \; \gamma_{3}\; a(x).
\end{align}
The variation of modified NC- action is 
\begin{align}
\delta{S_{F}}&=-\int d^{2}x\; \big[ \overline{\psi}(x)\star\; \gamma^{\mu}\gamma_{3}\; \partial_{\mu}a(x)\star \; \psi(x)+ &\\& \nonumber
\overline{\eta}(x)\star \; \gamma^{\mu}\gamma_{3}\; \partial_{\mu}a(x)\star \; \eta(x)+2iM \; \overline{\eta}(x)\; \gamma_{3}\star \; a(x)\star \; \eta(x) \big].
\end{align}
By virtue of cyclic property associated with $\star$-product \eqref{cyclic}, we can write 
\begin{equation}
\int d^{2}x \big[ \overline{\psi}(x)\star \; \gamma^{\mu}\gamma_{3}\; \partial_{\mu}a(x)\star \; \psi(x)+ \overline{\eta}(x)\star \; \gamma^{\mu}\gamma_{3}\; \partial_{\mu}a(x)\star \; \eta(x) \big]= \int d^{2}x \; a(x)\star \; \partial_{\mu}J^{\mu}_{3}(x).
\end{equation}
The axial Ward identity can be computed by imposing the following condition
\begin{equation}
\frac{\delta{W}}{\delta{a(x)}} \big{\vert}_{a(x)=0}=0 ,
\end{equation}
Which is equivalent to 
\begin{equation}
\partial_{\mu}<J^{\mu}_{3}(x)>=\lim _{M\rightarrow \infty} 2iM \; <\overline{\eta}(x)\; \gamma_{3}\star \; \eta(x)> . 
\end{equation}
The averaged regularized chiral current in two dimensions is 
\begin{align}
<J^{\mu}_{3}(x)>=\\ \nonumber &\frac{\int \mathcal{D}\psi \mathcal{D}\overline{\psi}\mathcal{D}\eta \mathcal{D}\overline{\eta} (\overline{\psi}\gamma^{\mu}\gamma_{3}\star \psi+\overline{\eta}\gamma^{\mu}\gamma_{3}\star \eta) e^{-\int d^{2}x \overline{\psi}\star i\slashed{D}\star \psi -\int d^{2}x\overline{\eta}\star (i\slashed{D}-M)\star \eta} }{\int \mathcal{D}\psi \mathcal{D}\overline{\psi}\mathcal{D}\eta \mathcal{D}\overline{\eta} e^{-\int d^{2}x \overline{\psi}\star i\slashed{D}\star \psi -\int d^{2}x\overline{\eta}\star (i\slashed{D}-M)\star \eta} },
\end{align}
Where $\eta$, $\overline{\eta}$ are the regulator and antiregulator Bose-like fields respectively.\\ The anomaly in two dimensions 
is 
\begin{equation} \label{anomaly2}
<\overline{\eta}(x)\; \gamma_{3}\;\star  \eta(x)>=\frac{\int \mathcal{D}\eta \; \mathcal{D}\overline{\eta}\; (\overline{\eta}(x)\; \gamma_{3}\; \star \eta(x))\; \star e^{-\int d^{2}x\; \overline{\eta}(i\slashed{D}-M)\star \;\eta}}{\int \mathcal{D}\eta \; \mathcal{D}\overline{\eta}\; e^{-\int d^{2}x\; \overline{\eta}\star (i\slashed{D}-M)\star \;\eta}}
\end{equation}
The  eigenvalues of Dirac operator are defined as 
\begin{equation}
\slashed{D}\star \phi_{n}(x)=\lambda_{n}\phi_{n}(x)
\end{equation}
Where $\phi_{n}$ are the eigenfunctions that correspond to the eigenvalue $\lambda_{n}$.  They  satisfy the following  orthonormal conditions 
\begin{align}
\sum \phi^{\dagger}_{n}(x)\star \; \phi(y)=\delta^{4}(x-y), \\ \nonumber
\int d^{4}x \; \phi^{\dagger}_{n}(x)\star \; \phi_{m}(x)=\delta_{nm}.
\end{align}
We expand the Pauli-Villars fields using the eigenfunctions of Dirac operator as a basis 
\begin{align}
\eta(x)=\sum \; c_{n}\; \phi_{n}(x), \\ \nonumber
\overline{\eta}(x)=\sum \; \overline{c}_{n}\; \phi^{\dagger}_{n}(x).
\end{align}
Where $c$ and $\overline{c}$ are ordinary complex numbers not a Grassmann numbers since the Pauli-Villars field $\eta$ and $\overline{\eta}$ obey Bose statistics. 
We insert the previous field expansion in  the equation \eqref{anomaly2},  we get  
\begin{equation}
<\overline{\eta}(x)\; \gamma_{3}\star \;   \eta(x)>=\frac{\prod_{n}\; \int dc_{n}\; \int d\overline{c}_{n}\; \sum_{\ell,m}\; \overline{c}_{\ell}\; \phi^{\dagger}_{\ell}(x)\; \gamma_{3}\; c_{m}\star \; \phi(x)_{m}\; e^{-\sum_{k}\; \overline{c}_{k}\; c_{k}(i\lambda_{k}-M)}}{\prod_{n}\int dc_{n}\int d\overline{c}_{n}\; e^{-\sum_{k}\; \overline{c}_{k}\; c_{k}\; (i\lambda_{k}-M)}}
\end{equation}
After performing the integrals, we find the simple expression for the anomaly 
\begin{equation}
<\overline{\eta}(x)\; \gamma_{3}\star \; \eta(x)>=\sum_{\ell}\; \frac{\phi^{\dagger}_{\ell}(x)\; \gamma_{3}\; \star \phi_{\ell}(x)}{i\; \lambda_{\ell}-M}. 
\end{equation}
The sum can be rewritten as  $\sum_{\ell}\; \phi^{\dagger}_{\ell}(x)\; \gamma_{3}\; \star \phi_{\ell}(x)=\mathrm{Tr}{\gamma_{3}\; \delta{(0)}}$ since $\gamma_{3}$ is a tracless matrix 
and $\delta(0)$ is infinite. Note that  the previous sum is  not well defined, so we need to regularize it 
in order to get a finite expression at the end of  calculations.  One way to regularize it is by  using Dirac operators,
\begin{equation}
\lim_{M\rightarrow \infty}\; 2iM\; <\overline{\eta}(x) \; \gamma_{3}\star \; \eta(x)>=\lim_{M\rightarrow \infty}\; 2iM \; \Tr \big[ \gamma_{3} \langle x | \frac{1}{i\slashed{D}-M}|x \rangle \big]
\end{equation}
After we multiply the numerator and denominator of the previous equation by $-i\slashed{D}-M$, we get 
\begin{align}
\lim_{M\rightarrow \infty}\; 2iM\; <\overline{\eta}(x)\; \gamma_{3}\star  \; \eta(x)>=\\ \nonumber \lim_{M\rightarrow \infty} 2M \; \Tr \big[ \gamma_{3} \langle x|   \frac{\slashed{D}}{\slashed{D}^{2}+M^{2}}| x \rangle \big]-\lim_{M \rightarrow \infty} 2iM^{2}\; \Tr \big[ \gamma_{3}\; \langle x | \frac{1}{\slashed{D}^{2}+M^{2}}|x \rangle \big]
\end{align}

We perform a Dyson expansion of the term $ \frac{1}{\slashed{D}^{2}+M^{2}}$ 
\begin{align}
\lim_{M\rightarrow \infty}\; 2iM\; <\overline{\eta}(x)\; \gamma_{3}\star  \; \eta(x)> \\ \nonumber&= -\lim_{M\rightarrow \infty} 2iM^{2}\; \Tr\int  \frac{d^{2}k}{(2\pi)^{2}}\;\frac{ \big[ \gamma_{3}\;\; g\;  \sigma_{\mu\nu}\;\; F^{\mu\nu}(x)_{\star} \big]}{2(k^{2}+M^{2})^{2}}\\ \nonumber
&= \frac{-g\epsilon^{\mu\nu}}{2\pi}\;F_{\mu\nu}(x)_{\star}
\end{align}
Where we have used the trace properties of $\gamma$ matrices in two dimensions 
and taking the limit when $M\rightarrow \infty$. 

We have used the following  identity during the previous calculations 
\begin{align}
\slashed{D}^{2}=D_{\mu}\gamma^{\mu}\; \star \; D_{\nu}\gamma^{\nu}=\frac{1}{4} \big( [\gamma^{\mu},\gamma^{\nu}]+\{\gamma^{\mu},\gamma^{\nu}\}\big)\big(\{D_{\mu},D_{\nu}\}_{\star}+[D_{\mu},D_{\nu}]_{\star}\big)
&\\ \nonumber 
=D_{\mu}\star D^{\mu}+\frac{g}{2}\; \sigma_{\mu\nu} \; F^{\mu\nu}_{\star}
\end{align}
Where $\sigma_{\mu\nu}=\frac{i}{2}[\gamma_{\mu},\gamma_{\nu}]$ and  $F_{\mu\nu}(x)_{\star}$ is the two-form non-commutative field strength.

Finally, the divergence of $J^{\mu}_{3}$ is given by 
\begin{equation}
\partial_{\mu}J^{\mu}_{3}=-\frac{g \epsilon_{\mu\nu}}{2\pi}F^{\mu\nu}(x)_{\star }
\end{equation}

\subsection{Four-Dimensional Anomaly}
The modified NC -massive Dirac action in four dimensions  is 
\begin{equation}
S_{F}=\int d^{4}x\; [ \overline{\psi}(x)\star\; (i\gamma^{\mu}D_{\mu}-m)\; \psi(x)+ \overline{\eta}(x)\star \; (i\gamma^{\mu}D_{\mu}-M)\; \eta(x)] .
\end{equation}
Spinors act in a Complex vector space $\mathbb{V}$ endowed  with a certain algebraic structure. The bases of spinor vector space satisfy the anti-commuting relation 
$ \{\gamma^{\mu},\gamma^{\nu}\}=2\;g^{\mu\nu}$, which is a Clifford algebra $\mathcal{C}\ell(1,3)$. 
$\gamma$ matrices are defined as Endomorphisms $ End(\mathbb{V}) $ of the complex vector space $\mathbb{V}$.  
\\We have considered the Dirac representation for $\gamma$ matrices  
\begin{align}
\gamma^0 = \begin{pmatrix} 
1 & 0 & 0 & 0 \\
0 & 1 & 0 & 0 \\ 
0 & 0 & -1 & 0 \\
0 & 0 & 0 & -1 \end{pmatrix} \quad
\gamma^1 = \begin{pmatrix}
0 & 0 & 0 & 1 \\
0 & 0 & 1 & 0 \\
0 & -1 & 0 & 0 \\
-1 & 0 & 0 & 0 \end{pmatrix}  \\ \nonumber
\gamma^2 = \begin{pmatrix}
0 & 0 & 0 & -i \\
0 & 0 & i & 0 \\
0 & i & 0 & 0 \\
-i & 0 & 0 & 0 \end{pmatrix} \quad 
\gamma^3 = \begin{pmatrix}
0 & 0 & 1 & 0 \\
0 & 0 & 0 & -1 \\
-1 & 0 & 0 & 0 \\
0 & 1 & 0 & 0 \end{pmatrix}
\end{align}
and the chirality is defined as  
\begin{equation}
\gamma_{5}=i\;\frac{ \epsilon^{\mu\nu\rho\sigma}}{4!}\; \gamma_{\mu}\; \gamma_{\nu}\; \gamma_{\rho}\; \gamma_{\sigma}=\;i \gamma^{0}\; \gamma^{1}\; \gamma^{2}\; \gamma^{3} ,
\end{equation}
Where $\epsilon^{\mu\nu\rho\sigma}$ is totally  antisymmetric tensor. 

The trace formulas in four dimensions are
\begin{equation}
\Tr(\gamma_{5}\; \gamma_{\mu}\;\gamma_{\nu})=o
\end{equation}
\begin{equation}
\Tr(\gamma_{5}\; \gamma_{\mu}\; \gamma_{\nu}\; \gamma_{\rho}\; \gamma_{\sigma})=-4i\; \epsilon_{\mu\nu\rho\sigma}
\end{equation}
\begin{equation}
\Tr(\gamma_{\mu}\gamma_{\nu})=4\; g_{\mu\nu}
\end{equation}
alongside with the fact that trace of any odd number of $\gamma$ matrices is zero. \\
The Euclidean effective action is 

\begin{equation}
e^{-W}=\int \mathcal{D}\psi \; \mathcal{D}\overline{\psi}\; \mathcal{D}\eta\; \mathcal{D}\overline{\eta}\; e^{-S_{F}}
\end{equation}
and the corresponding  conserved vector current and axial-vector current are respectively 
\begin{equation}
J_{\mu}=\overline{\psi}(x)\star\; \gamma_{\mu}\; \psi(x)+\overline{\eta}(x)\star\; \gamma_{\mu}\; \eta(x)
\end{equation}

\begin{equation}
J^{5}_{\mu}=\overline{\psi}(x)\star\; \gamma_{\mu}\gamma^{5}\; \psi(x)+\overline{\eta}(x)\star\; \gamma_{\mu}\gamma^{5}\; \eta(x)
\end{equation}
Under chiral rotation and up to first order in $a(x)$, the matter fields transform like 
\begin{align}
\psi^{'}(x)\rightarrow  \psi(x)+i\; \gamma_{5}\; a(x)\; \star \psi(x)
, \\ \nonumber
\overline{\psi^{'}}(x)\rightarrow \overline{\psi}(x)+i\; \overline{\psi}(x)\star \; \gamma_{5}\; a(x), \\ \nonumber
\eta^{'}(x)\rightarrow \eta(x)+i\; \gamma_{5}\; a(x)\star \; \eta(x), \\ \nonumber
\overline{\eta^{'}}(x)\rightarrow \overline{\eta}(x)+i\; \overline{\eta}(x)\star \; \gamma_{5}\; a(x). 
\end{align}
The variation of modified Pauli-Villars action is 
\begin{align}
\delta{S_{F}}&=-\int d^{4}x\; \big[ \overline{\psi}(x)\star\; \gamma^{\mu}\gamma_{5}\; \partial_{\mu}a(x)\star \; \psi(x)+ 2im \; \overline{\psi}(x)\; \gamma_{5}\star \; a(x)\star \; \psi(x)+ 
&\\& \nonumber \overline{\eta}(x)\star \; \gamma^{\mu}\gamma_{5}\; \partial_{\mu}a(x)\star \; \eta(x)+2iM \; \overline{\eta}(x)\; \gamma_{5}\star \; a(x)\star \; \eta(x) \big]
\end{align}
Using the cyclic property of $\star$-product \eqref{cyclic}
\begin{equation}
\int d^{4}x \big[ \overline{\psi}(x)\star \; \gamma^{\mu}\gamma_{5}\; \partial_{\mu}a(x)\star \; \psi(x)+ \overline{\eta}(x)\star \; \gamma^{\mu}\gamma_{5}\; \partial_{\mu}a(x)\star \; \eta(x) \big]= \int d^{4}x \; a(x)\star \; \partial_{\mu}J^{\mu}_{5}(x)
\end{equation}
The following condition gives the so-called axial-Ward identity 
\begin{equation}
\frac{\delta{W}}{\delta{a(x)}} \big{\vert}_{a(x)=0}=0
\end{equation}
The axial-Ward (AW) identity is 
\begin{equation}
\partial_{\mu}<J^{\mu}_{5}(x)>=2im\;<\overline{\psi}(x) \; \gamma_{5}\star \; \psi(x)>+ \lim _{M\rightarrow \infty} 2iM \; <\overline{\eta}(x)\; \gamma_{5}\star \; \eta(x)>
\end{equation}
The real field $ \psi$  obeys Fermi statistics and the ghost field( regulator) $\eta$ obeys a Bose statistics. Thus,the 
regulator is used only for fermionic loops and that is what we need since our aim is  to calculate the axial anomaly. \\
The second term in the right-hand side of previous equation is 
\begin{equation} \label{anomaly}
<\overline{\eta}(x)\; \gamma_{5}\;\star  \eta(x)>=\frac{\int \mathcal{D}\eta \; \mathcal{D}\overline{\eta}\; (\overline{\eta}(x)\; \gamma_{5}\; \star \eta(x))\; \star e^{-\int d^{4}x\; \overline{\eta}(i\slashed{D}-M)\star \;\eta}}{\int \mathcal{D}\eta \; \mathcal{D}\overline{\eta}\; e^{-\int d^{4}x\; \overline{\eta}(i\slashed{D}-M)\star \;\eta}}
\end{equation}
We expand the fields in orthonormal basis like the case of two-dimensional anomaly 
\begin{align}
\eta(x)=\sum \; c_{n}\; \phi_{n}(x), \\ \nonumber
\overline{\eta}(x)=\sum \; \overline{c}_{n}\; \phi^{\dagger}_{n}(x). 
\end{align}
We insert the expansion in equation \eqref{anomaly}
\begin{align}
<\overline{\eta}(x)\; \gamma_{5}\star \;   \eta(x)>=\\ \nonumber &\frac{\prod_{n}\; \int dc_{n}\; \int d\overline{c}_{n}\; \sum_{\ell,m}\; \overline{c}_{\ell}\; \phi^{\dagger}_{\ell}(x)\; \gamma_{5}\; c_{m}\star \; \phi(x)_{m}\; e^{-\sum_{k}\; \overline{c}_{k}\; c_{k}(i\lambda_{k}-M)}}{\prod_{n}\int dc_{n}\int d\overline{c}_{n}\; e^{-\sum_{k}\; \overline{c}_{k}\; c_{k}\; (i\lambda_{k}-M)}}
\end{align}
After performing the integrals in previous equation, we obtain 
\begin{equation}
<\overline{\eta}(x)\; \gamma_{5}\star \; \eta(x)>=\sum_{\ell}\; \frac{\phi^{\dagger}_{\ell}(x)\; \gamma_{5}\; \star \phi_{\ell}(x)}{i\; \lambda_{\ell}-M}
\end{equation}
As we have seen in the previous section, the $\sum_{\ell}\; \phi^{\dagger}_{\ell}(x)\; \gamma_{5}\; \star \phi_{\ell}(x)$ is  not well defined. We solve this problem in the  virtue of Dirac operators 
\begin{equation}
\lim_{M\rightarrow \infty}\; 2iM\; <\overline{\eta}(x) \; \gamma_{5}\star \; \eta(x)>=\lim_{M\rightarrow \infty}\; 2iM \; \Tr \big[ \gamma_{5} \langle x | \frac{1}{i\slashed{D}-M}|x \rangle \big]
\end{equation}
We multiply the numerator and denominator of the previous equation by $-i\slashed{D}-M$. This gives
\begin{align}
\lim_{M\rightarrow \infty}\; 2iM\; <\overline{\eta}(x)\; \gamma_{5}\star  \; \eta(x)>&\\ \nonumber =\lim_{M\rightarrow \infty} 2M \; \Tr \big[ \gamma_{5} \langle x|   \frac{\slashed{D}}{\slashed{D}^{2}+M^{2}}| x \rangle \big]-\lim_{M \rightarrow \infty} 2iM^{2}\; \Tr \big[ \gamma_{5}\; \langle x | \frac{1}{\slashed{D}^{2}+M^{2}}|x \rangle \big]
\end{align}

By performing Dyson expansion of the term $ \frac{1}{\slashed{D}^{2}+M^{2}}$ and due to the trace properties of $\gamma$ matrices in four dimensions 
and at the limit when $M\rightarrow \infty$. The axial anomaly is 
\begin{align}
\lim_{M\rightarrow \infty}\; 2iM\; <\overline{\eta}(x)\; \gamma_{5}\star  \; \eta(x)> & \\ \nonumber= -\lim_{M\rightarrow \infty} 2iM^{2}\; \Tr\;\int  \frac{d^{4}k}{(2\pi)^{4}}\frac{ \big[ \gamma_{5}\;\; g^{2}\sigma_{\mu\nu}\; \sigma_{\rho\sigma}\; F^{\mu\nu}(x)\star \; F^{\rho\sigma}(x) \big]}{4(k^{2}+M^{2})^{3}} & \\  \nonumber
= \frac{-g^{2}\epsilon^{\mu\nu\rho\sigma}}{16\pi^{2}}\; \Tr\; (F_{\mu\nu}(x)\star \; F_{\rho\sigma}(x))
\end{align}
The previous result matches the result in \cite{Sadooghi}. \\
Finally, the divergence of $J^{\mu}_{5}$ is given by 
\begin{equation}
\partial_{\mu}J^{\mu}_{5}=2im\;  \Tr [\gamma_{5}<x| \frac{1}{i\slashed{D}-m}| x>]- \frac{g^{2}\epsilon^{\mu\nu\rho\sigma}}{16\pi^{2}}\; \Tr\; (F_{\mu\nu}(x)\star \; F_{\rho\sigma}(x))
\end{equation}
\section{Conclusion}
We have calculated the $U(1)$  axial anomaly in two and four dimensions using the modified path integral formalism coupled to Pauli-Villars regulators  in the  presence of noncommutative background. One advantage of this method is that  we can easily convert from the non-perturbative path integral formalism to the perturbative axial Ward identities at each step in the derivation. This  resilience is maintained in this method while in other methods such transition can not be established. We also expect that our developed method can be used to calculate anomalies for any gauge group not only for $U(1)$. Also, it can be used in calculating chiral anomalies in curved spacetimes. We hope that our study will give evidence about the power of path integral formalism in comparison with other perturbative methods.

\appendix
\section{Fujikawa Method In Four-Dimensions} 
We will follow the same general procedures that is performed in \cite{Sadooghi} with Gaussian function regularization.\\
The partition function in NC-QED is 
\begin{equation}
\mathcal{Z}= \int \mathcal{D}\psi \; \mathcal{D}\overline{\psi}\; e^{-iS_{F}}.
\end{equation}
The chiral transformations for the matter fields $\psi$ and $\overline{\psi}$ are 
\begin{align}
\psi \rightarrow \psi^{'}(x)=U^{5}(x)\star\;  \psi(x), \\ \nonumber
\overline{\psi}(x) \rightarrow \overline{\psi^{'}}(x)=\overline{\psi}(x)\star\; U^{5}(x). 
\end{align}
Where $U^{5}(x)=(e^{i\gamma_{5}\; a(x)})_{\star}$.\\
The NC-massless Dirac action is 
\begin{equation}
S_{F}[\psi,\overline{\psi}]=i\; \int d^{4}x\; \overline{\psi}(x)\; \gamma^{\mu}\; \star \partial_{\mu}\psi(x). 
\end{equation}
under chiral rotation the previous action transforms like 
\begin{equation} 
S_{F}\rightarrow  S_{F}^{'}=S_{F}-\int d^{4}x\; \overline{\psi}(x)\star \; \partial_{\mu}a(x)\; \gamma^{\mu}\gamma^{5}\star \psi(x) 
\end{equation}
From the cyclic property of  $\star$-Product \eqref{cyclic}, we can write 
\begin{equation}
\int d^{4}x\; \overline{\psi} (x) \star \partial_{\mu}a(x)\; \gamma^{\mu}\gamma^{5}\; \star \psi(x)= i\;\int d^{4}x\; a(x)\; \star \partial_{\mu}J^{\mu}_{5}(x)
\end{equation}
Where $J^{\mu}_{5}(x)$ is the axial-current. From  previous property, the massless Dirac action becomes 
\begin{equation}
S^{'}_{F}=S_{F}-i\int d^{4}x\; a(x)\star \; \partial_{\mu}J^{\mu}_{5}(x).
\end{equation}
The  Partition function under Chiral rotaion reads as 
\begin{align}
\mathcal{Z}^{'}&=\int \mathcal{D}\psi^{'}\; \mathcal{D}\overline{\psi}(x)^{'}\; e^{-iS^{'}_{F}}\\ \nonumber &= \int \mathcal{D}\psi^{'}\; \mathcal{D}\overline{\psi}^{'}\; e^{-iS_{F}-\int d^{4}x\; a(x)\star \; \partial_{\mu}J^{\mu}_{5}(x)}
\end{align}
Under chiral transformations, Dirac fields transforms like 
\begin{align}
\overline{\psi}^{'}(x)\approx \overline{\psi}(x)+i\; \gamma_{5}\; \overline{\psi}(x)\star a(x), \\ \nonumber
\psi^{'}(x) \approx  \psi(x)+ i\; \gamma_{5}\; a(x)\; \star \psi(x).
\end{align}
The fermionic measure changes like
\begin{align}
\mathcal{D}\psi^{'}\; \mathcal{D}\overline{\psi}^{'}&=(\mathcal{J})^{-2}\; \mathcal{D}\psi\; \mathcal{D} \overline{\psi}\\ \nonumber
&= e^{i\int d^{4}x\; a(x)\star \; \mathcal{A}(x)}\; \mathcal{D}\psi\; \mathcal{D} \overline{\psi}
\end{align}
Where $\mathcal{J}$ is the Jacobian. \\ 
The sum of exponential is not well defined $\sum_{n}^{\infty}\; \phi^{\dagger}_{n}(x)\; \gamma_{5}\star \phi(x)_{n}=\mathrm{Tr} \gamma_{5}\; \delta(0)$.   This is  because $\gamma_{5}$ is a traceless matrix and   $\delta(0)$ is infinite. Thus, we 
need to regularize it by introducing a Gaussian cutoff  
\begin{equation}
\sum_{n}\; \phi^{\dagger}_{n}(x)\; \gamma_{5}\star \phi_{n}(x)=\lim _{M\rightarrow \infty} \sum_{n}\; \phi^{\dagger}_{n}(x)\; \star e^{-\frac{D_{\mu}\; \gamma^{\mu}\; \star D_{\nu}\; \gamma^{\nu}}{M^{2}}}\; \star \phi_{n}(x) 
\end{equation}
Where $M$ is a large mass and  Dirac delta function is defined in the same way as in the commutative case. \\
It is convenient to perform  Fourier transformations $ \phi_{n}(x)=\int \frac{d^4k}{(2\pi)^{2}}\; e^{ik.x}\; \widetilde{\phi}_{n}(k)$ of fields.  The Gaussian sum can be expressed as 
\begin{equation}\label{Gaussian}
\sum_{n}\; \phi^{\dagger}_{n}(x)\star \; \gamma_{5}\phi_{n}(x)= \lim _{M\rightarrow \infty} \int \frac{d^{4}k}{(2\pi)^{4}}\; \mathrm{Tr}\; e^{-ik.x}\; \gamma_{5}\; e^{-\frac{D_{\mu}\; \gamma^{\mu}\; \star D_{\nu}\; \gamma^{\nu}}{M^{2}}}\; e^{ik.x},
\end{equation}
where the trace in  previous relation is taken over the $\gamma$ matrices and over the Lie algebra of the  action. \\
We can do the following algebraic manipulation of Dirac operators 
\begin{align}
-D_{\mu}\gamma^{\mu}\; \star \; D_{\nu}\gamma^{\nu}=-\frac{1}{4} \big( [\gamma^{\mu},\gamma^{\nu}]+\{\gamma^{\mu},\gamma^{\nu}\}\big)\big(\{D_{\mu},D_{\nu}\}_{\star}+[D_{\mu},D_{\nu}]_{\star}\big)
\\ \nonumber =-D_{\mu}\star D^{\mu}-\frac{g}{2}\; \sigma_{\mu\nu}\; F^{\mu\nu}
\end{align}
where the two-form field strength is defined as $ ig\; F_{\mu\nu}=[D_{\mu},D_{\nu}]_{\star}$ and $ \sigma_{\mu\nu}=\frac{i}{2}\; [\gamma_{\mu},\gamma_{\nu}]$. 
We insert the previous identity inside the Gaussian sum \ref{Gaussian}. Also,  we perform a scaling of momenta $ k_{\mu}\rightarrow M k^{'}_{\mu}$, 
note that the differential Dirac operators will make a plane wave shift.  Finally we arrive to 
\begin{align}
\sum_{n}\; \phi^{\dagger}_{n}(x)\star \; \gamma_{5}\; \phi_{n}(x)\\ &\nonumber= \lim_{M \rightarrow \infty}\; \int \frac{M^{4}d^{4}k^{'}}{(2\pi)^{4}}\; e^{k^{2}}\; \mathrm{Tr}\big[ \gamma_{5}\; e^{-\frac{(D_{mu}+iM\; k^{'}_{\mu})\star (D^{\mu}+iM\;k^{'\mu})}{M^{2}}}\;\star e^{-\frac{i\gamma_{\mu}\; \gamma_{\nu}F^{\mu\nu}}{2M^{2}}} \big]
\end{align}

We expand the exponents in previous relation. By using the $\gamma$ trace properties with $M\rightarrow \infty$ gives the singlet anomaly  
\begin{equation}
\mathcal{A}(x)=\frac{i\; g^{2}\epsilon^{\mu\nu\rho\sigma}}{16\pi^{2}}\; \mathrm{Tr}(F_{\mu\nu}(x)\star \; F_{\rho\sigma}(x)), 
\end{equation}
and the divergence of  chiral current is 
\begin{equation}
\partial_{\mu}J^{\mu}_5(x)=\frac{-g^{2}\epsilon^{\mu\nu\rho\sigma}}{16\pi^{2}}\; \mathrm{Tr}\; (F_{\mu\nu}(x)\star \; F_{\rho\sigma}(x)). 
\end{equation}


\begin{thebibliography}{99}
	
	\bibitem{Connes}
	A. Connes, M. R. Douglas, and A. Schwarz\emph{ Noncommutative Geometry and Matrix Theory: Compactification on Tori}, \emph{JHEP} {\bf 9802:003} (1998).
	
	\bibitem{Seiberg}
	N. Seiberg and E. Witten, \emph{Noncommutative Geometry And String Theory}, \emph{JHEP}
	{\bf 9909:032} (1999).
	
	
	
	\bibitem{Minwalla}
	S.Minwalla, M. Van Raamsdonk, and N. Seiberg, \emph{Noncommutative Perturbative Dynamics},
	\emph{JHEP} {\bf 0001:028} (2000).
	\bibitem{Ardalan}
	F.Ardalan, H.Arfaei and  M.M.Sheikh-Jabbari, \emph{Noncommutative geometry from strings and branes}, \emph{JHEP} {\bf 9902:016} (1999).
	\bibitem{Lledo}
	S. Ferrara and M.A. Lledo, \emph{Some Aspects of Deformations of Supersymmetric Field Theories}, \emph{JHEP} {\bf  0005} (2000).
	
	\bibitem{Sadooghi}
	F.Ardalan and  N.Sadooghi, \emph{Axial Anomaly In Noncommutative QED on $R^{4}$}, \emph{Int.J.Mod.Phys.}, {\bf A16} (2001). 
	\bibitem{Martin}
	J. M. Gracia-Bondia and C. P. Martin , \emph{Chiral Gauge Anomalies on Noncommutative $\mathbb{R}^{4}$}, \emph{Phys.Lett.B}, {\bf479} (2000).
	
	.
	
	
	
	\bibitem{Chams}
	A.H.Chamseddine, \emph{$SL(2,\mathbb{C})$ Gravity with Complex Vierbein and Its Noncommutative Extension}, \emph{Phys.Rev. D} {\bf69 } (2004).
	
	\bibitem{Fujikawa}
	K.Fujikawa, \emph{Generalized Pauli-Villars Regularization and the Covariant Form of Anomalies}, \emph{Nucl.Phys. B}{\bf 428} (1994).
	\bibitem{Banerjee}
	R.Banerjee, \emph{Anomalie and the Index Theorem In The Path Integral approach}, \emph{HD-THEP} {\bf 25} Heidelberg University (1990).
	
\end{thebibliography}
\end{document}